\documentclass{mn2e}

\usepackage{graphicx,times,array}

\def \kms {${\rm{km}\,\rm{s}^{-1}}$}

\def \xsh  {X-SHOOTER}
\def \egx  {ESO325--G004}

\def \ezgal  {{\sc EzGal}}

\def \nai {Na{\,\sc i}}
 \def \rev  {}

\def \apjs{ApJS}

\def \apj{ApJ}
\def \aj{AJ}
\def \mnras{MNRAS}

 \title[A lightweight giant elliptical galaxy]{A giant elliptical galaxy with a lightweight initial mass function\thanks{Based on observations collected at the European Southern Observatory, Chile 
(ESO Programmes 077.A-0806(A), 088.B-0653(C) and 291.B-5011(A)).}\thanks{{\rev
 Based on observations made with the NASA/ESA Hubble Space Telescope, obtained at the Space Telescope Science Institute, which is operated by the Association of Universities for Research in Astronomy, Inc., under NASA contract NAS 5-26555. These observations are associated with programs 10429 and 10710.}
}}
\author[Russell J. Smith \& John R. Lucey]
{Russell J. Smith\thanks{Email: russell.smith@durham.ac.uk} and 
John R. Lucey
~\\
Department of Physics, University of Durham, Durham DH1 3LE\\
}

\date{Accepted 2013 June 20. Received 2013 June 17; in original form 2013 May 21}

\pagerange{\pageref{firstpage}$-$\pageref{lastpage}} \pubyear{2013}

\begin{document}

\label{firstpage}

\maketitle

\begin{abstract}
We present new observations of the closest-known strong-lensing galaxy, 
the $\sigma$\,$\approx$\,330\,\kms\ giant elliptical ESO325--G004, made with the ESO Very Large Telescope.
The low redshift  of the lens ($z_{\rm l}$\,=\,0.035) results in arcs being formed
at a small fraction of the effective radius, ($R_{\rm Ein}$\,=\,2.85\,arcsec\,$\approx$\,$R_{\rm eff}/4$). 
At such small radii, stars dominate the lensing mass, so that lensing provides a direct probe of the stellar mass-to-light ratio, with only
small corrections needed for dark matter. However, the redshift of the galaxy lensed by \egx\ was unknown until now, so the lensing mass was not securely determined.
Using X-SHOOTER, we have detected multiple spectral lines, from two bright parts of the arc system, and measured a source redshift
of  $z_{\rm s}$\,=\,2.141.
Combined with lens modelling constraints, this yields a total mass inside the Einstein radius of 1.50$\pm$0.06$\times10^{11}M_\odot$.
We estimate the range of possible contribution of dark matter to the lensing mass, using halo profile statistics from cosmological N-body simulations.
Subtracting this component yields a stellar mass-to-light ratio for the lens of $M_*/L_{\rm F814W}$\,=$\,3.14^{+0.24}_{-0.42} (M/L)_{\odot, \rm F814W}$. 
Using VIMOS, we have also obtained very high signal-to-noise spectroscopy for the lens galaxy. Fitting models to this spectrum confirms
that  \egx\ has a very old stellar population. 
For a Milky-Way-like (Kroupa) IMF, the stellar population fit yields a predicted stellar mass-to-light ratio of $\Upsilon_{\rm MW}$\,=\,3.01$\pm$0.25$(M/L)_{\odot, \rm F814W}$.
Hence the mass attributable to stars with a Kroupa IMF is consistent with the lensing estimate. 
By contrast, a Salpeter (or heavier) IMF is disfavoured at the 99.8\,per cent confidence level.
A ``heavyweight'' IMF, with a mass twice as large as the Kroupa case, is firmly excluded for this galaxy. 
Such an IMF has been proposed for more distant elliptical lenses, 
and also to explain strong dwarf-star sensitive spectral features, in particular the Na{\,\sc i} 8200\,\AA\ doublet.
A FORS2 far-red spectrum shows that this feature is as strong in \egx\ as it is in other high-$\sigma$ ellipticals, 
suggesting tension between dwarf-star indicators and lensing-mass constraints for this galaxy. 
\end{abstract}   

\renewcommand{\textfraction}{0.05}

\section{Introduction}\label{sec:intro}

The distribution of stellar masses at formation (the initial mass function,
IMF) is a crucial quantity in astrophysics, both as a constraint on star-formation
processes and in linking observed luminosities to the stellar masses of galaxies. 
It is therefore of great importance to establish whether the IMF is universal or, if not, how it depends systematically  
on the environment in which stars form.

Within the Milky Way and its satellites, the IMF can be determined directly through star counts. The distribution
follows a power law with the Salpeter (1955) slope (${\rm d}N(M)$\,$\propto$\,$M^{-x}{\rm d}M$ with $x$\,=\,2.35)
for $M$\,$\ga$0.5\,$M_\odot$, but breaks to a shallower slope at lower mass (e.g. Kroupa 2001; Chabrier 2003). There is little
evidence for systematic variation in IMF as a function of metallicity, star-formation rate or other properties in the Milky Way itself (Bastian, Covey \& Meyer 2010).
However, IMFs with slopes flatter than Salpeter (at $\sim$0.7\,$M_\odot$) 
have been reported for several dwarf satellites, which probe to lower metallicities 
(Wyse et al. 2002; Kalirai et al. 2013; Geha et al. 2013). 

For galaxies beyond the Milky Way and its immediate neighbours, resolved star counts are impossible, and indirect methods are used.
In this case, the mass-to-light ratio of the stellar population  ($M_*/L$) provides a simple constraint on the IMF. For a single power law, slopes 
steeper than the Salpeter $x$\,=\,2.35 imply large numbers of very faint dwarf stars which dominate the mass; for much flatter slopes, the mass
budget becomes dominated by stellar remnants. In either case, $M_*/L$ is increased relative to the Salpeter power-law.
Breaking the power law away from Salpeter at low mass, as observed in the Milky Way, yields lower $M_*/L$ than a single power-law, by $\sim$35\,per cent.
Combining rotation curves with stellar population models for a sample of spiral galaxies, Bell \& de Jong (2001) found that a $\sim$30\,per cent
reduction in mass, relative to Salpeter, was required to avoid violating dynamical constraints on the total mass.
Hence a Milky-Way-like (Chabrier or Kroupa) IMF appears to be generic for spiral galaxies as a class. 

For elliptical galaxies, constraining the the IMF via $M_*/L$  poses a greater challenge, since masses are more difficult to establish
for dynamically-hot systems. 
Strong gravitational lensing of background galaxies provides a powerful method to determine masses in these objects. 
Important progress has been made through the systematic assembly and follow up of large samples of lenses, especially from the 
Sloan Lens ACS (SLACS) survey (Bolton et al. 2006). In the SLACS methodology, lenses are selected through the presence of anomalous emission lines in the 
galaxy spectrum due to the lensed source, and followed up with {\it Hubble Space Telescope} (HST) imaging. 
Modelling the lensing configuration provides the total projected mass within an aperture, while the velocity dispersion from SDSS spectroscopy yields an 
additional dynamical constraint,
which allows the stellar contribution to be decoupled from the dark-matter halo. 
Analysing 56 SLACS lenses, Treu et al. (2010, hereafter T10) found that for a universal standard Navarro, Frenk \& White (1996, NFW) halo, some 80\,per cent
of the total lensing mass was contributed by the ``stellar'' model component. Comparing the lensing stellar mass against the mass determined from stellar
population fits to broadband colours, T10 found that Salpeter IMFs were favoured over Milky-Way-like distributions on average, and that the mass normalisation
of the IMF increases with galaxy velocity dispersion. For the most massive SLACS galaxies, with $\sigma>300$\,\kms, the analysis requires an IMF 
twice as heavy as the Kroupa IMF. 
{\rev Recent lensing analysis of massive spiral galaxies suggests there are variations within such galaxies, with bulges having heavier IMFs than disks (Dutton et al. 2013).}
Dynamical modelling estimates for nearby ellipticals also indicate a larger $M_*/L$ than expected for a Milky-Way IMF (e.g. Thomas et al. 2011;
Wegner et al. 2012; Cappellari et al. 2013).
{\rev In the largest of these studies (Cappellari et al.), the average excess for the most massive galaxies is compatible with Salpeter IMF, rather than the more
extreme forms required by SLACS.}
Note however, that the dynamical studies include few galaxies with very high velocity dispersion $\sigma>300$\,\kms.

As  noted above, large $M_*/L$ ratios could arise either from an excess of faint dwarf stars in a ``bottom-heavy'' IMF, or from an excess of dark remnants in 
a ``top-heavy''  IMF. The analysis of gravity-sensitive spectroscopic absorption features promises to distinguish between these cases, 
by isolating lines and bands characteristic of either dwarf or giant stars (Spinrad \& Taylor 1971;  Whitford 1977; Cohen 1978; Faber \& French 1980; 
Carter, Visvanathan \& Pickles 1986; Couture \& Hardy 1993; Cenarro et al. 2003; Conroy \& van Dokkum 2012a, hereafter CvD12a).
This method, updated with modern spectral synthesis model ingredients, was applied to a small sample of massive ellipticals by van Dokkum \& Conroy (2010), who 
found strong dwarf-star features which could only be reproduced in models with a very bottom-heavy IMF. 
Following this work, a number of studies have confirmed an apparent excess of low-mass stars in massive ellipticals, compared to the Milky Way.
(Spiniello et al. 2012; Conroy \& van Dokkum 2012b, hereafter CvD12b; Smith, Lucey \& Carter 2012; Ferreras et al. 2013; La Barbera et al. 2013; Spiniello et al. 2013). 
The degree of dwarf-star-enrichment, and the strength of its dependence on galaxy mass, metallicity and other properties, is still not fully clear however. 
In general, analyses which include the \nai\ 8200\,\AA\ doublet feature have tended to find stronger evidence for dwarf enrichment, 
a discrepancy already noted by Carter et al. (1986), and persisting to the latest works (e.g. figure 12 of CvD12b).
A particular challenge is to decouple the IMF effect from trends in abundance ratios, especially Na/Fe which affects not only the \nai\ doublet
but also many other lines, through its strong influence on the free electron pressure in cool stellar atmospheres (CvD12a).
For the most massive ellipticals ($\sigma>300$\,\kms), CvD12b favour IMFs with mass normalisation twice that of the Milky Way IMF, 
in close concordance with the SLACS lensing results.

In summary, several recent studies have presented evidence for ``heavyweight'' IMFs\footnote{We use the 
term ``heavyweight'' to refer to the high mass normalisation, without reference to whether this arises from dwarf stars or from remnants.}
in giant ellipticals, with 
a mass-to-light  ratio twice that of a 
Milky-Way-like IMF. 
Given the important and widespread implications of this result, careful observational scrutiny is essential. 
In this paper, we exploit an unusual low-redshift lens system to measure the stellar 
mass-to-light ratio in a single, but very powerful, $\sigma$\,$>$\,300\,\kms\ elliptical galaxy.
In Smith et al. (2005, hereafter S05), we discovered  a system of gravitationally-lensed arcs
around ESO325--G004, using HST imaging. 
This was a serendipitous discovery, in the sense that it was not derived from any systematic search for lenses. 
Due to the closeness of this lens ($z_{\rm l}$\,=\,0.035), the Einstein radius in \egx\ is smaller
than the stellar effective radius, by a factor of four. 
Hence in this system the lensing mass is dominated by stars to an unusual degree, and only small corrections
for dark matter are required.
However, the lensing mass has not been determined until now, 
because the redshift of the background source was unknown. 
In this paper, we report the measurement of the source redshift and 
determine the implications for the stellar mass-to-light ratio and IMF in \egx. 

The structure of the paper is as follows.
Section~\ref{sec:obs} presents the observations, including 
	measurement of the source redshift (Section~\ref{sec:xshoot}), 
	photometry and determination of the total lensing mass  (Section~\ref{sec:hst}), 
	and spectroscopy of the lens galaxy to determine its age, and hence the mass-to-light ratio expected for a given IMF (Section~\ref{sec:vimos}), 
In Section~\ref{sec:dmcor}, we estimate the likely contribution of dark matter to the lensing mass. 
Section~\ref{sec:alphapost} compares the dark-matter-corrected lensing mass against the age constraints
to infer the viable range of IMF normalization, and presents tests for the robustness of our analysis. 
In Section~\ref{sec:disc} we compare our results to those obtained from SLACS, and to the results from dwarf-star indicators,
including a measurement of the \nai\ 8200\,\AA\ feature for \egx\ itself.
Brief conclusions are summarized in Section~\ref{sec:concs}.

Where necessary, we adopt cosmological parameters from WMAP7: 
{\rev $H_0$\,=\,70.4\,\kms\,Mpc$^{-1}$, $\Omega_{\rm m}$\,=\,0.272 and $\Omega_{\rm \Lambda}$\,=\,0.728 
(Komatsu et al. 2011).}

\begin{figure}
\includegraphics[angle=0,width=85mm]{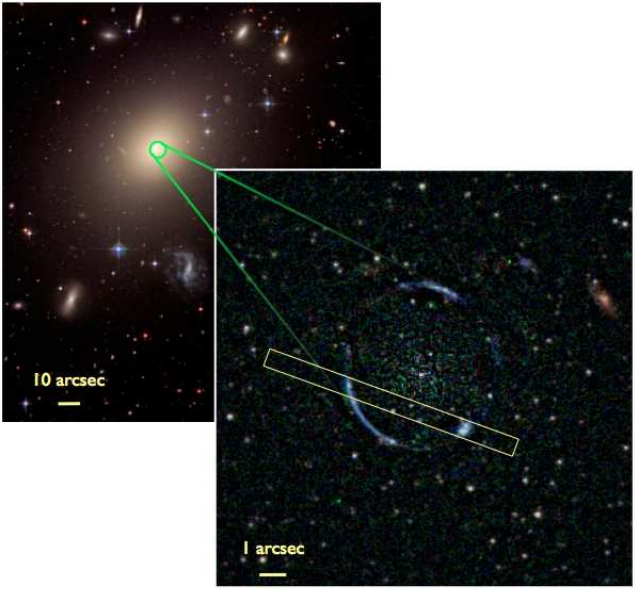}
\vskip -3mm
\caption{
HST image of the $z$\,=\,0.035 giant elliptical galaxy ESO325--G004 (Smith et al. 2005) and its immediate environment
(Credit: NASA/ESA and the Hubble Heritage Team).
The inset shows a zoom of the central regions of the image after subtracting a smooth model describing the lens galaxy.
The arcs are formed at the Einstein radius of 2.85\,arcsec (1.96\,kpc at the distance of the lens). 
Because this is small compared to the effective radius (12.3\,arcsec), the enclosed mass is dominated by stars, rather than dark matter {\rev (see Section~\ref{sec:dmcor})}.
The colour figures were created from F475W (blue, 4800\,s exposure), F625W (green, 2400\,s) and F814W (red, 18900\,s) images taken with the
Advanced Camera for Surveys. The yellow rectangle indicates the slit orientation for the \xsh\ observations.
}
\label{fig:slit}
\end{figure}

\section{Observations}\label{sec:obs}

\subsection{\xsh\ spectroscopy: arc redshift}\label{sec:xshoot}

We observed \egx\  with the \xsh\  three-arm echelle spectrograph (Vernet et al. 2011), mounted on UT2 of the ESO Very Large Telescope (VLT),  on 2013 March 7.
Spectra were obtained with a 0.4\,arcsec slit, providing resolving power 10000, 18000 and 10500 in the UVB, VIS and NIR arms respectively. 
The total integration time was 2400\,s, split between two exposures.
The image quality, as estimated from the acquisition frames, was $\sim$0.5\,arcsec FWHM. The slit was aligned to intersect
two segments of the arc system (Arc C and the brightest part of Arc A, in the nomenclature of S05), as shown in Figure~\ref{fig:slit}.

\begin{figure*}
\includegraphics[angle=0,width=175mm]{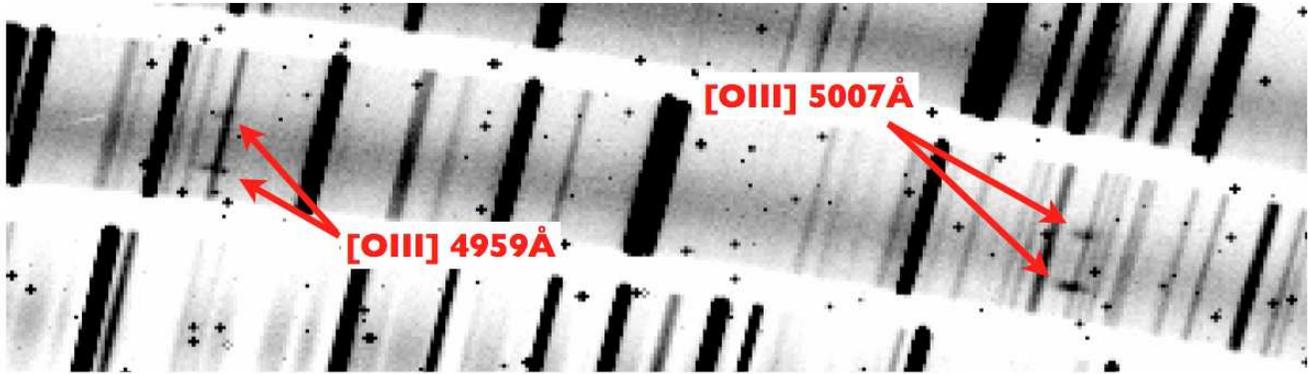}
\vskip -5mm
\caption{Extract from the raw, two-dimensional spectrum from a single \xsh\ NIR exposure, showing the redshifted [O\,{\sc iii}]  lines from the lensed source. 
The slit was aligned to intersect two arcs (see Figure~\ref{fig:slit}); lines from both are clearly visible, on either side of the diffuse trace of the lens-galaxy continuum.
}
\label{fig:twodspec}
\end{figure*}

\begin{figure*}
\includegraphics[angle=0,width=175mm]{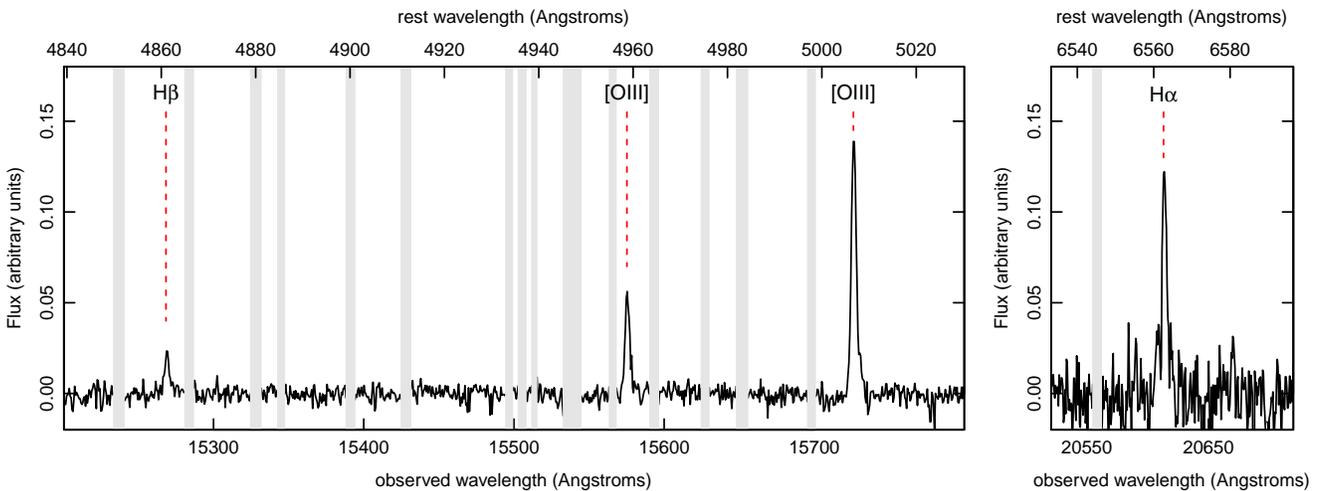}
\caption{Segments of the one-dimensional spectrum of the lensed source.
The signal has been combined over both arcs and both exposures. Areas affected by strong sky lines are masked in grey. 
A smooth background (including any true continuum) has been subtracted from the spectrum.
Emission line positions are shown for a redshift of $z_{\rm s}$\,=\,2.141.}
\label{fig:onedspec}
\end{figure*}

Visual inspection of the raw data revealed the presence of emission lines only in the NIR arm spectra.
Identical line emission is observed from the two arc segments, confirming beyond reasonable doubt that the source is indeed a multiply-imaged lensed galaxy
(Figure~\ref{fig:twodspec}).
The observations were reduced using the standard \xsh\ pipeline, to produce a rectified and wavelength-calibrated two-dimensional spectrum. An approximate correction
for telluric absorption was applied using a standard star observation.
One-dimensional sky-subtracted spectra were extracted centred on each arc and combined to yield the final spectrum,
extracts from which are shown in Figure~\ref{fig:onedspec}. Despite the short total integration time and the small number of exposures (hence poor rejection of cosmetic defects),
four emission lines are  detected, at wavelengths corresponding to  [O\,{\sc iii}]  $\lambda\lambda$ 4594, 5007\,\AA, H$\beta$ and H$\alpha$ for a source 
redshift of $z_{\rm s}$\,=\,2.141. 
The characteristics of the lensed source are not the concern of this paper, but we note the spectrum is similar to those of other lensed high-redshift 
star-forming galaxies (Richard et al. 2011).

\subsection{HST Photometry and lensing analysis}\label{sec:hst}

\egx\ was originally observed with the Advanced Camera for Surveys on HST in January 2005 (programme 10429, P.I. Blakeslee),
for 18900\,s in F814W and 1100\,s in F475W, as reported by S05. 
Deeper observations in blue passbands were obtained in February 2006 (programme 10710, P.I. Noll) for a Hubble Heritage public release, providing
4800\,s in F475W and 2400\,s in F625W.

The lensing mass was determined by S05, modulo the then-unknown source redshift, using a singular isothermal sphere (SIS) model
with an additional external shear term. From this model, S05 found an Einstein radius of $R_{\rm Ein}$\,=\,2.85\,arcsec and a corresponding mass
(projected within  $R_{\rm Ein}$)  of $M_{\rm Ein}^{\rm SIS}$\,=\,$1.40\times10^{11}  (D_{\rm s}/D_{\rm ls}) M_\odot$. 
Here, $D_{\rm s}$ is the angular-diameter distance from the observer to the source 
and  $D_{\rm ls}$ is the angular-diameter distance from the lens to the source.
For the measured source redshift $z_{\rm src}$\,=$\,2.141$, the geometric factor 
is close to unity, $D_{\rm s}/D_{\rm ls}$\,=\,1.027 (with negligible error).
Hence the lensing mass for the SIS model is $M_{\rm Ein}^{\rm SIS}$\,=\,$1.44\times10^{11}\,M_\odot$.
The effective (half-light) radius of \egx, determined from the F814W image is $R_{\rm eff}$\,=\,12.3$\pm$0.5\,arcsec, a factor of four larger than $R_{\rm Ein}$.

We derive the luminosity projected within the Einstein radius from simple aperture photometry performed on the HST/ACS in F814W and F475W 
filters\footnote{The transformation to Johnson--Cousins magnitudes applied in S05 was erroneous, leading to a mass-to-light ratio substantially lower than we report here.}.
We work entirely in the native photometric bandpasses, at the observed redshift of \egx, and express all magnitudes in the Vega system.
The observed aperture magnitudes are F814W\,=\,13.543 and F475W\,=\,15.568. Extinction corrections are 
$A_{\rm F814W}$\,=\,$0.092$ and $A_{\rm F475W}$\,=\,0.196, from Schlafly \& Finkbeiner (2011). 
We assume a luminosity distance of 152\,Mpc  (WMAP7 cosmology, no peculiar velocity), and hence a distance modulus of 35.909\,mag. 
The absolute magnitude of the Sun, redshifted to $z$\,=\,0.034 in the observed bands, is 4.066 in F814W and 5.254 in F475W 
(determined using \ezgal, Mancone \& Gonzalez 2012).
Hence the luminosities are 
$L_{\rm F814W}$\,=\,$4.07\times10^{10} L_{\odot, \rm F814W}$ and 
$L_{\rm F475W}$\,=\,$2.07\times10^{10} L_{\odot, \rm F475W}$. 
We adopt a 2\,per cent error on luminosity to account for absolute calibration uncertainties (the statistical errors are much smaller).
The ratio of SIS lensing mass to luminosity gives the total $M_{\rm Ein}^{\rm SIS}/L_{\rm F814W}$\,=\,3.54$\pm$0.06\,$(M/L)_\odot$.

Since the \egx\ lensing mass is expected to be dominated by stellar mass, rather than by dark matter {\rev (see Section~\ref{sec:dmcor}}), we have also modelled the lensing configuration
using a mass distribution proportional to the observed luminosity. 
To fit this mass-follows-light (MFL) model, we treat the lens as a set of point masses and compute the net deflection experienced by image-plane 
pixels corresponding to the arcs (as identified in the deep F475W image). For the ``mass'' image, we use a smooth model derived from ellipse 
fitting to the F814W image, and incorporating harmonic terms to describe the slightly boxy isophote shape. 
The lens model is then specified by the (total) mass-to-light ratio $M/L$, 
{\rev plus a linear shear term, with free amplitude and direction,
intended to account for additional distortions due to nearby structures.}
Given values for these parameters, we determine the source-plane location of the arc pixels, and their 
likelihood of being drawn from a single compact region on the source plane. The assumed intrinsic source is a circular Gaussian with 0.35\,arcsec FWHM. 
Interpreting this likelihood as the probability that the lens model is correct, we use a 
Markov Chain Monte Carlo method to sample from the probability distribution of the model parameters. 
Marginalizing over the shear amplitude and direction, this method yields $M_{\rm Ein}^{\rm MFL}/L_{\rm F814W}$\,=\,$3.69\pm0.06$, 
marginally larger than the SIS result. 
Hereafter, we adopt the results of the MFL for the lensing mass-to-light ratio, and the nominal Einstein radius is that derived from the SIS model.
{\rev Thus the mass within $R_{\rm Ein}$ is $M_{\rm Ein}^{\rm MFL}$\,=\,$1.50\times10^{11} M_\odot$, i.e. 4\,per cent larger than found from the (over-simplistic) SIS model.
Other lens models which account for the angular structure of the luminous matter (e.g. a singular isothermal ellipse, with or without external shear) yield similar $M_{\rm Ein}$ 
to the MFL approach, within $\sim$1\,per cent. The robustness of $M_{\rm Ein}$, with respect to reasonable choices for the mass model,
is a standard result in lensing studies (e.g. Kochanek 1991; Koopmans et al. 2006; Treu 2010). In principle, 
$M_{\rm Ein}$ includes contributions from all structures along the line-of-sight to the source; in ${\rm \Lambda}$CDM cosmology, the rms contribution from large-scale 
structure is calculated to be $\sim$2\,per cent for a $z$\,=\,2 source (Taruya et al. 2002). 

In what follows, we adopt the lensing mass from the MFL model, $M_{\rm Ein}$\,=\,$1.50\pm0.06\times10^{11} M_\odot$.
The adopted 4\,per cent error reflects an conservative estimate of the systematic uncertainties, based on the difference between SIS- and MFL-model masses.
}

\subsection{VIMOS spectroscopy: lens properties}\label{sec:vimos}

The lensing $M/L$ yields information on the IMF if other parameters of the stellar population, in particular its age, can be constrained using additional data.

We observed \egx\ with VIMOS (Le Fevre et al. 2003) in integral-field unit (IFU) mode, on UT3 of the VLT in 2006 April--May.
The data were obtained using the (``old'') HR--blue grism, with a wavelength range of 4200--6200\,\AA\ and 
resolution of 1.65\,\AA\ FWHM, sampled at 0.54\,\AA\,pixel$^{-1}$. The spatial coverage was 13$\times$13\,arcsec$^2$, with a 
scale of 0.33\,arcsec\ per IFU fibre.
Eight individual spectra were obtained, each with integration time of 1865\,s, with pointing adjustments of a few arcsec between 
observations to average over fibre sensitivity variations.
The pipeline-reduced spectra from IFU elements within $R_{\rm Ein}$\,=\,2.85\,arcsec of the galaxy centre 
were combined from each observation separately to allow assessment of systematic errors between exposures. 
The overall signal-to-noise ratio for the full 4-hour integration is $\sim$400\,\AA$^{-1}$ at 5000\,\AA.

Standard Lick absorption indices were measured on the combined spectra and corrected to the standard Lick resolution ($\sim$9\,\AA\ FWHM, but 
wavelength dependent) and to  zero velocity dispersion, following the method described in Smith, Lucey \& Hudson (2007). 
The velocity dispersion measured from the extracted spectrum, and used for the resolution correction, is $\sigma$\,=\,335\,\kms.
Corrections from the flux-calibrated system to the Lick flux response system were applied using the offsets tabulated by Norris, Sharples \& Kuntschner (2006).

As may be expected given the very high signal-to-noise ratio, the scatter in index value between observations (e.g. rms 0.08\,\AA\ for H$\beta$) exceeds the formal error
on each individual observation (typically 0.04\,\AA\ for H$\beta$). The source of excess scatter appears to be slight ripples in the relative continuum shapes between
the observations. To account for the systematic error floor, we adopt the mean over the eight measurements and use the observed scatter to derive the 
error  in the mean (0.03\,\AA\ for H$\beta$). 
The spectrum obtained for \egx\ in the 6dF Galaxy Survey (Jones et al. 2004, 2009), in an aperture of radius 3.35\,arcsec yields index values compatible with the 
VIMOS measurements, but with uncertainties around six times larger.

We use the index data to derive constraints on the stellar mass-to-light ratio assuming a MW-like (Kroupa 2001) IMF.
We denote this quantity $\Upsilon_{\rm MW}$, while the true stellar mass-to-light ratio is $M_*/L$, 
and $\alpha_{\rm MW}$\,=\,$(M_*/L)/\Upsilon_{\rm MW}$ is the ``mass normalisation factor'' of the true IMF relative to Kroupa.
In this convention, a Chabrier IMF has $\alpha_{\rm MW}$\,=\,0.87, a Salpeter IMF has $\alpha_{\rm MW}=1.55$,  and a ``heavyweight" IMF as favoured by
SLACS and CvD12b for massive ellipticals has $\alpha_{\rm MW}$\,=\,2.

To determine $\Upsilon_{\rm MW}$, we work in the context of models by Maraston (2005) and 
Thomas et al. (2003, 2004), loosely referred to collectively as M05 hereafter. 
The M05 model set has the advantage of incorporating $\alpha$-element enhancements in the index predictions (though not explicitly in the broadband fluxes), as well as
covering a comfortable range in super-solar total metallicity. 
We assume single-burst star-formation history models throughout the analysis. This simplification can be justified on the grounds that the galaxy 
shows no evidence for {\it recent} star formation, and that an extended star-formation history at {\it early} epochs is indistinguishable in terms of 
indices and colours from a single-burst. Inspection of the 6dF red-arm spectrum does not show any evidence for emission at H$\alpha$, and 
hence there is no reason to suspect emission infilling contamination of the H$\beta$ and H$\gamma$ lines.

For a fixed IMF, the stellar mass-to-light ratio of a population depends mainly on age, and to a lesser extent on metallicity. Since individual line-strength indices 
depend on age, metallicity and abundance ratios (especially $\alpha$/Fe), multiple indices are needed to constrain the age.
We use H$\beta$ and H$\gamma_{\rm F}$ as the primary age indicators, together with the $\alpha$-element-dominated Mg$b$ index and three iron-tracing 
indices (Fe5015, Fe5270, Fe5335). 
We compute the likelihood of the index data for all six features
at each point in a grid spaced uniformly in log(age) (from log(5\,Gyr) to log(12.5\,Gyr)), total metallicity [Z/H] (from 0.1 to 0.4\,dex) 
and [$\alpha$/Fe] (from 0.1 to 0.5\,dex). 
We determine the  $\Upsilon_{\rm MW}$ at each point in the age--metallicity grid using 
the \ezgal\ code of Mancone \& Gonzalez (2012) to compute the mass-to-light ratio in the observed-frame F814W band, in solar units, consistent
with the convention used in our HST photometry. 
The M05 models assume no variation in $\Upsilon_{\rm MW}$ with $\alpha$/Fe, but this should be a small effect. For example,  Percival et al. (2008)
find that old $\alpha$-enhanced ([$\alpha$/Fe]\,=\,+0.4) stellar population models are 2--3\,per cent brighter in the I-band 
than models with solar-scaled abundances at the same age and total metallicity.

\begin{table}
\caption{Line-strength indices, as measured from the VIMOS spectrum, expressed in Angstroms. 
Values in the third column have been corrected for velocity broadening and corrected to the
Lick-system resolution and flux-response system. Errors were derived from the scatter among eight
separate observations.
}\label{tab:ixvals}
\center
\begin{tabular}{lcc}
\hline
Index & raw &  corrected \\
\hline

H$\beta$ & \phantom{--}$1.68\pm0.03$ &  \phantom{--}$1.63\pm0.03$ \\ 
H$\gamma_{\rm F}$ & --$2.09\pm0.08$    &   --$2.08\pm0.08$ \\ 
Mg$b$ & \phantom{--}$4.22\pm0.04$ & \phantom{--}$4.90\pm0.05$  \\ 
Fe5015  & \phantom{--}$4.18\pm0.10$ &  \phantom{--}$5.29\pm0.12$ \\
Fe5270  & \phantom{--}$2.46\pm0.05$ &  \phantom{--}$2.92\pm0.06$  \\ 
Fe5335 & \phantom{--}$1.95\pm0.04$ & \phantom{--}$2.83\pm0.05$ \\
\hline
\end{tabular}
\vskip 4mm
\end{table}

\begin{figure}
\includegraphics[angle=0,width=85mm]{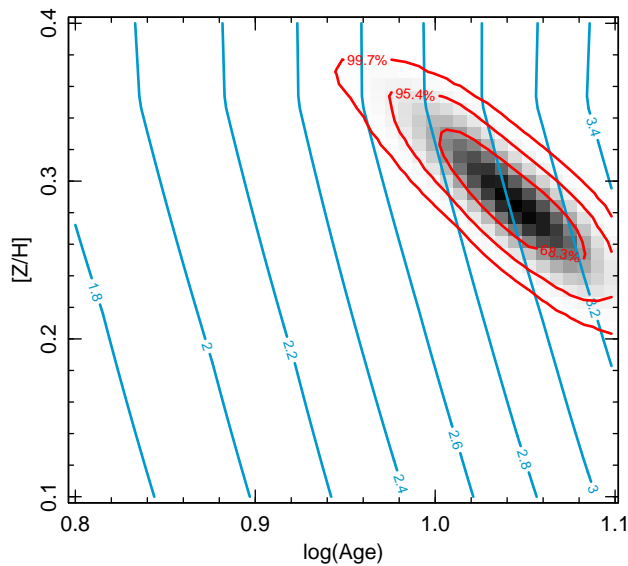}
\caption{
Determination of $\Upsilon_{\rm MW}$, i.e. the stellar mass-to-light ratio for a MW-like IMF, from spectroscopy of the lens. The grey-scale and
red contours show the probability distribution (marginalized over [$\alpha$/Fe]) for the age and metallicity, derived from fits to the measured indices. 
The blue contours indicate the mass-to-light ratio (in F814W, labelled in solar units) corresponding to each location on the grid. 
}
\label{fig:agecons}
\end{figure}

\begin{figure*}
\includegraphics[angle=270,width=175mm]{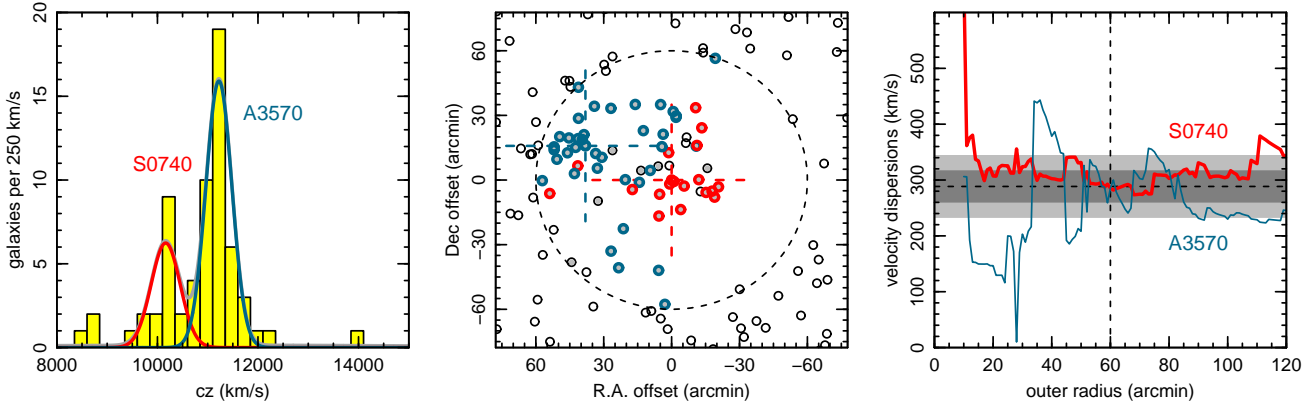}
\caption{The velocity dispersion of Abell S0740. Left: Histogram of velocity measurements from the NASA Extragalactic Database for galaxies within 60\,arcmin of ESO325--G004.
The curve shows the best-fit Gaussian mixture model, highlighting the components for Abell S0740 (red) and Abell 3570 (blue). Centre: Sky distribution of the galaxies with velocity
measurements. For galaxies within  60\,arcmin of ESO325--G004, membership assignments are indicated with the same colours as in the left-hand panel.
Grey points with a black outline indicate the galaxies which were assigned to the smooth background. Unfilled symbols mark galaxies which were
excluded before fitting.
The assignment algorithm makes no use of the spatial information, but galaxies assigned to Abell S0740 and to Abell 3570 are clearly centred near their respective dominant members
(indicated by the cross-hairs). Right: the velocity dispersions of the two components derived from the Gaussian mixture model, 
as a function of the cut-off radius (the adopted value of 60\,arcmin is marked by the vertical line). The 1$\sigma$ and 2$\sigma$ error regions for Abell S0740 are indicated by the
dark and light grey bands respectively.}
\end{figure*}

\begin{figure*}
\includegraphics[angle=0,width=175mm]{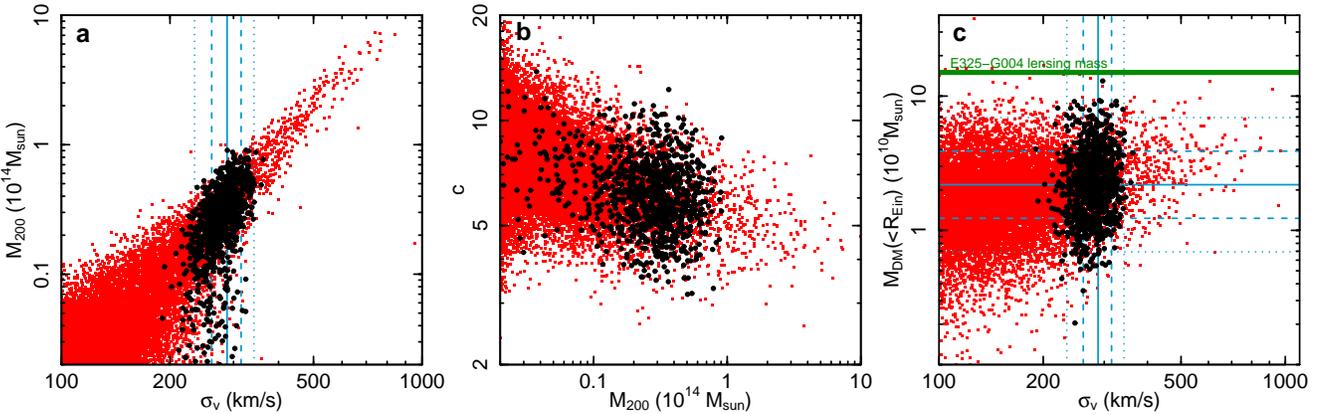}
\caption{Constraints on the dark-matter contribution to the lensing mass derived from halos in the Millennium Simulation. Left: the relationship between 
virial mass and velocity dispersion (the latter estimated from semi-analytic member galaxies, for approximate match to the observational methods). The vertical bands show the 
measured velocity dispersion for Abell S0740 (the group which is dominated by \egx). The red points show a uniform sampling of all halos while the black points are sampled with 
probability according to the measured velocity dispersion. Centre: the NFW concentration parameters assigned following the mass--concentration relation of Neto et al. (2007),
as a function of virial mass. Right: the dark-matter mass projected inside the Einstein radius, computed for an NFW profile with the assigned mass and concentration, as a function
of velocity dispersion. The horizontal lines show the median and 68- and 95-percent intervals for enclosed dark-matter mass, which is well-fit by a Gaussian in $\log M_{\rm DM}$. 
The thick green line shows the total lensing mass $M_{\rm Ein}$.
}\label{fig:enchalomass}
\end{figure*}

Figure~\ref{fig:agecons} shows how the index data constrain the mass-to-light ratio.
We derive the final probability distribution for $\Upsilon_{\rm MW}$  by weighting the predictions according to their likelihoods (implicitly 
marginalizing over all the stellar population parameters), which yields $\Upsilon_{\rm MW}$\,=\,3.01$\pm$0.14\,$(M/L)_{\odot, \rm F814W}$.
As a test of systematics within this method, we re-ran the analysis excluding each index in turn from the constraint set. 
As may be expected, the Balmer indices have the largest effect on the derived mass-to-light ratio. 
If H$\gamma{\rm_F}$ is excluded, we obtain  $\Upsilon_{\rm MW}$\,=\,2.70$\pm$0.18\,$(M/L)_{\odot, \rm F814W}$, while
If H$\beta$ is excluded, we recover  $\Upsilon_{\rm MW}$\,=\,$3.25^{+0.06}_{-0.11}$\,$(M/L)_{\odot, \rm F814W}$. 
(The asymmetric errors arise from imposing a hard upper bound on the age, i.e. the galaxy
is not permitted to be older than the Universe in the adopted cosmology.)
Since these results differ by more than the error in the fit with both Balmer lines, 
we inflate the error to account for systematic uncertainties. 
The final estimated stellar population mass-to-light ratio, under the assumption of a Kroupa IMF, is 
 $\Upsilon_{\rm MW}$\,=\,3.01$\pm$0.27\,$(M/L)_{\odot, \rm F814W}$.
An equivalent analysis for the F475W band yields $\Upsilon_{\rm MW}$\,=\,6.32$\pm$0.63\,$(M/L)_{\odot, \rm F475W}$.

To verify the robustness of our results, we have also applied a full-spectrum fitting method to the VIMOS data, using the CvD12a models. 
Within this model set, the spectra are best matched (and well matched) by models with the maximum age of 13.5\,Gyr. Although the 
derived age is larger than from the M05 index-fitting approach,
this is compensated by slightly smaller mass-to-light ratios at given age in CvD12a.  In fact, 
the best-fitting model has $\Upsilon_{\rm MW}$\,=\,2.97\,$(M/L)_{\odot, \rm F814W}$
(after converting from Chabrier to Kroupa IMF, to match our convention). Hence the results from the two approaches, using different 
models and different fitting methods, are indistinguishable.

We note also that assuming a single-burst population has little impact on the $\Upsilon_{\rm MW}$ derived for the F814W band.
To illustrate this, consider a two-burst star-formation history. The impact of the younger burst is {\it smaller} in $\Upsilon_{\rm MW}$ 
than on the age derived from the V-band. Hence for a {\it fixed} V-luminosity-weighted age (e.g. 9\,Gyr), a two-burst model
(e.g. 96\,per cent 12\,Gyr, 4\,per cent 2\,Gyr, by mass) has slightly {\it larger}  $\Upsilon_{\rm MW}$
(3.25\,$(M/L)_{\odot, \rm F814W}$) than a single burst  (2.85\,$(M/L)_{\odot, \rm F814W}$).

In summary, analysing the VIMOS spectrum confirms that the stellar population of \egx\ is very old, and hence
has a high mass-to-light ratio for a given IMF. Even for an Kroupa IMF, stars alone contribute a mass
$\Upsilon_{\rm MW} L_{\rm Ein}$\,=\,1.2$\pm$0.1\,$\times$10$^{11}\,M_\odot$ within the Einstein radius, which is
80$\pm$7\,per cent of the total lensing mass.

\section{Dark matter contribution}\label{sec:dmcor}

The lensing mass refers to the total mass projected within the Einstein radius, including both stellar mass (living stars and remnants) and dark matter\footnote{
In principle, there are also contributions from gas and from a central supermassive black hole.
We assume the gas mass projected within the Einstein radius is negligible. This is equivalent to assuming
all gas is initially converted into stars, and that the gas lost in winds and supernovae is either recycled into further generations
of stars or else expelled into a hot, low-density halo. Black hole mass contributions are small, and addressed in Section~\ref{sec:alphapost}.}.
In this section, we use the statistics of halo profiles in a cosmological N-body simulation to estimate the 
dark-matter correction and, crucially, the uncertainty in the correction.

To help constrain the dark-matter contribution in \egx, we use the velocity dispersion of its surrounding halo to select appropriate
halos from the simulation. The lens is the dominant member of a small galaxy group, catalogued Abell S0740 (Abell, Corwin \& Olowin 1989),
which is located close to another system, Abell 3570, at a projected distance of $\sim$40\,arcmin and similar redshift.
To measure the velocity dispersion of  Abell S0740, we use the available (incomplete) redshift information compiled from the 
NASA Extragalactic Database\footnote{The majority of these redshifts are derived from the 6dF Galaxy Survey (Jones et al. 2004, 2009).}.
Selecting galaxies within a radius of 60\,arcmin and $|\Delta{}cz|<5000$\,\kms\ from ESO325--G004, we fit the redshift distribution using a Gaussian mixture model. 
The model includes two components representing 
Abell S0740 and Abell 3750, with mean redshift fixed to the velocities of their dominant galaxies (10164\,\kms\ and 11223\,\kms\ respectively), but fitting for the 
velocity dispersions. We allow an additional broad component to describe an approximately uniform background distribution.
The best-fitting velocity dispersion for Abell S0740 under this model is $\sigma_v$\,=\,288$\pm$26\,\kms\ (the error is obtained by resampling
using the posterior classification probabilities). 
We confirm the robustness of this measurement by re-fitting the model varying the outer cut-off radius, 
finding that $\sigma_v$ is stable within the formal $2\sigma$ error
range for cut-offs of 30--110\,arcmin.

Neglecting (until Section~\ref{sec:alphapost}) the possible contraction of the dark-matter halo in response to the dense baryonic component (Blumenthal et al. 1986), 
we can use dark-matter-only simulations to estimate the contribution of the halo to the projected mass inside the Einstein radius. 
We first select halos from the Millennium Simulation (Springel et al. 2005) having virial masses $M_{200}$ greater than $10^{12}\,M_\odot$ and compute 
their line-of-sight velocity dispersions
based on member galaxies assigned in the semi-analytic model of De Lucia \& Blaizot (2007). 
We then draw a large random sample from these halos, 
with selection probability given by a Gaussian describing our constraint on $\sigma_v$ for Abell S0740, i.e. with mean 288\,\kms\ and standard deviation 26\,\kms. 
We represent each halo by an NFW profile (Navarro et al. 1996), with concentration parameter, $c$, assigned according to the 
statistical distribution determined as a function of mass  by Neto et al. (2007) for ``relaxed'' halos. We assume that at fixed halo mass, $c$ is 
is uncorrelated with the velocity dispersion, since the latter is obtained from all galaxies assigned to the halo, and hence is a large-aperture measurement. 
To compute $M_{\rm DM}$, the dark-matter mass projected within the Einstein radius, for a given halo mass and concentration,  we employ the analytic results 
presented by {\L}okas \& Mamon (2001). The relationships between $\sigma_v$, $M_{200}$, $c$ and $M_{\rm DM}$ are shown in Figure~\ref{fig:enchalomass}.
The derived distribution of dark halo contribution for the $\sigma_v$-matched sample can be accurately represented by a Gaussian in
 $\log(M_{\rm DM}/M_\odot)$, with mean $10.34$ and standard deviation 0.25. 
 Perhaps surprisingly, the distribution of  $M_{\rm DM}$ is only weakly dependent on velocity dispersion: although large $\sigma_v$ is a predictor for higher halo mass, 
 such halos have lower concentration and hence a smaller fraction of their mass projected inside $R_{\rm Ein}$. 
 
 Comparing this distribution to the lensing estimate, we find that dark matter contributes $15^{+11}_{-6}$\,per cent of the total mass projected within $R_{\rm Ein}$, 
 in the absence of  baryonic contraction effects. The estimated dark matter component, added to the stellar mass from Section~\ref{sec:vimos} (80$\pm7$\,per cent with a Kroupa IMF), 
 is thus sufficient to reproduce the observed lensing configuration.

\section{Constraints on the IMF}\label{sec:alphapost}

\begin{figure*}
\includegraphics[angle=0,width=165mm]{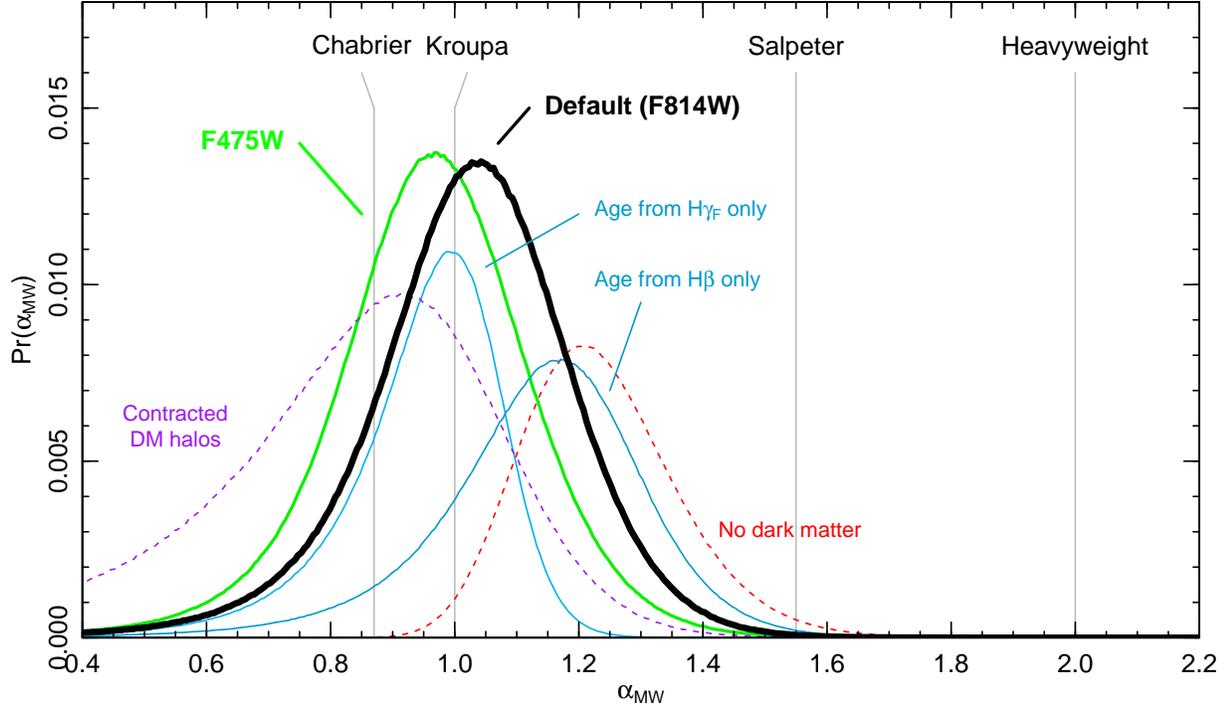}
\vskip -7mm
\caption{Probability distribution function for the IMF mass normalisation factor $\alpha_{\rm MW}$, marginalizing over age and metallicity, and the contribution from dark matter.
The thick black curve shows our default solution, while other curves indicate the effect of variations in the modelling. 
The green curve shows the results derived using the F475W data.
Blue curves illustrate the effect of using only one Balmer index (H$\beta$ or H$\gamma_{\rm F}$) in the fit. 
The red dashed line results from setting the dark matter contribution to be zero, while for the purple curve, all dark-matter contributions are doubled to indicate
the maximum likely halo contraction effect. Input parameters, best-fit $\alpha_{\rm MW}$ and the IMF probabilities associated with these curves are summarized 
in Table~\ref{tab:alphapost}. The mass normalization equivalent to Chabrier, Kroupa and Salpeter IMFs are indicated. The ``heavyweight'' line has 
$\alpha_{\rm MW}$\,=\,2, corresponding to the average for $\sigma$\,$>$\,300\,\kms\ ellipticals, as derived by T10 and CvD12b.}
\label{fig:alphapost}
\end{figure*}

\begin{table*}
\caption{Summary of constraints obtained for the IMF mass normalisation parameter $\alpha_{\rm MW}$ in \egx. The six lines correspond to the curves shown in 
Figure~\ref{fig:alphapost}. 
$\Upsilon_{\rm MW}$ is the mass-to-light ratio of the best-fitting stellar population model, assuming a Kroupa IMF; $L_{\rm Ein}$ is the luminosity inside the 
Einstein radius $R_{\rm Ein}$, $M_{\rm DM}$ is the estimated dark matter mass projected inside $R_{\rm Ein}$. 
Probabilities estimated from the distributions for $\alpha_{\rm MW}$ are:
	Pr$(\alpha_{\rm MW}$\,$<$\,$1)$, the probability that the IMF mass normalization is smaller than Kroupa; 
	Pr$(\alpha_{\rm MW}$\,$>$\,$1.55)$, the probability that the IMF is heavier than Salpeter; and 
	Pr$(\alpha_{\rm MW}$\,$>$\,$2)$, the probability that the IMF normalization is larger than 2.0, which is the mean $\alpha_{\rm MW}$ derived for $\sigma$\,$>$\,300\,\kms\ ellipticals
	by T10 and CvD12b.
}\label{tab:alphapost}
\setlength{\extrarowheight}{4pt}
\begin{tabular}{llccccccc}
\hline
&  & $\Upsilon_{\rm MW}/(M/L)_\odot$ & $L_{\rm Ein} / 10^{10}L_\odot$ & $\log M_{\rm DM}/M_\odot$ & $\alpha_{\rm MW}$ &  Pr($\alpha_{\rm MW}$\,$<$1) & Pr($\alpha_{\rm MW}$\,$>$1.55) & Pr($\alpha_{\rm MW}$\,$>$2)\\
\hline
(1) & Default                        & $3.01\pm0.27$ & 4.07$\pm$0.08 & $10.34\pm0.25$ & $1.04_{-0.15}^{+0.15}$ & $\phantom{<}$0.354686 & $\phantom{<}$0.001545 & $\phantom{<}$0.000002 \\
(2) & F475W band               & $6.32\pm0.63$ &  2.07$\pm$0.04 & $10.34\pm0.25$ & $0.97_{-0.15}^{+0.15}$ & $\phantom{<}$0.531162 & $\phantom{<}$0.000873 & $\phantom{<}$0.000002 \\
(3) & H$\gamma_{\rm F}$ only         & $3.25\,_{-0.11}^{+0.06}$ & 4.07$\pm$0.08 & $10.34\pm0.25$ & $0.99_{-0.11}^{+0.09}$ & $\phantom{<}$0.577770 & $<$0.000001 & $<$0.000001 \\
(4) & H$\beta$ only                  & $2.70\pm0.18$ & 4.07$\pm$0.08 &  $10.34\pm0.25$ & $1.17_{-0.15}^{+0.13}$ & $\phantom{<}$0.138487 & $\phantom{<}$0.006241 & $<$0.000001 \\
(5) & Contracted halos         & $3.01\pm0.27$ & 4.07$\pm$0.08 & $10.64\pm0.25$ & $0.91_{-0.21}^{+0.17}$ & $\phantom{<}$0.680350 & $\phantom{<}$0.000332 & $<$0.000001 \\
(6) & No dark matter                 & $3.01\pm0.27$ &4.07$\pm$0.08 & --- & $1.20_{-0.11}^{+0.13}$ & $\phantom{<}$0.012379 & $\phantom{<}$0.019302 & $\phantom{<}$0.000030 \\
\hline
\end{tabular}
\end{table*}

\begin{table*}
\caption{Summary of relevant parameters of the \egx\ lens system.}
\setlength{\extrarowheight}{4pt}
\begin{tabular}{llcl}
\hline
Quantity & symbol & value & comments\\
\hline
Lens redshift (heliocentric) && 0.0339 & From 6dF \\
Lens redshift (CMB frame) & $z_{\rm l}$ &  0.0347 & \\
Lens angular diameter distance & $D_{\rm l}$ & 142\,Mpc & From $z_{\rm l}$ with WMAP7 cosmology \\
Angular scale at lens &  & 0.687\,kpc/arcsec & From $D_{\rm l}$ \\
Lens half-light radius & $R_{\rm Eff}$ & 12.3\,arcsec & From ACS F814W image \\ 
Lens stellar velocity dispersion & $\sigma$ & 331$\pm$2\,\kms\ & $R_{\rm eff}$/8 aperture \\ 
Lens luminosity distance & & 152\,Mpc & From $z_{\rm l}$ with WMAP7 cosmology \\
Lens distance modulus &  & 35.909 \\
Source redshift & $z_{\rm s}$ & 2.141 &  From X-Shooter \\
Lensing geometry factor &  $f_z = D_{\rm s}/D_{\rm ls}$ & 1.027 & From $z_{\rm l}$ and $z_{\rm s}$ with WMAP7 cosmology \\
\hline
Lensing critical surface density &   $\Sigma_{\rm crit}$ & $5.70\times10^9\,M_\odot$\,arcsec$^{-2}$ & From $f_z$ and $D_{\rm l}$\\
Einstein radius & $R_{\rm Ein}$ & 2.85\,arcsec & Singular isothermal sphere mass model (S05) \\
Luminosity inside the Einstein radius & $L_{\rm F814W}$  & $4.07\pm0.08\times10^{10} L_{\odot, \rm F814W}$ & ACS photometry, corrected for extinction \\
Total lensing mass-to-light ratio & $M_{\rm MFL}/L_{\rm F814W}$ & $3.69\pm0.03  (M/L)_{\odot, \rm F814W}$ & From mass-follows-light model\\
Total lensing mass inside $R_{\rm Ein}$ & $M_{\rm Ein}^{\rm MFL}$ & $1.50\pm0.06 \times 10^{11} M_\odot$ & From mass-follows-light model\\
\hline
S0740 group velocity dispersion & $\sigma_v$ & $288\pm26$\,\kms & Literature redshifts; decomposed from Abell 3570\\
Dark matter mass within $R_{\rm Ein}$ & $M_{\rm DM}$ & $2.19^{+1.70}_{-0.96}\times10^{10} M_\odot$ & From Millennium simulation halo statistics\\
Stellar mass inside $R_{\rm Ein}$ & $M_*$ &  $1.28^{+0.10}_{-0.17}\times10^{11} M_\odot$ & From $M_{\rm Ein}^{\rm MFL}$ and $M_{\rm DM}$\\
Stellar mass-to-light ratio & $M_*/L_{\rm F814W}$ &  $3.14^{+0.24}_{-0.42} (M/L)_{\odot, \rm F814W}$ & From $M_*$ and $L_{\rm F814W}$\\
\hline
Stellar mass-to-light ratio for Kroupa IMF & $\Upsilon_{\rm MW}$  &  $3.01\pm0.25\,(M/L)_{\odot, \rm F814W}$  &  From fit to VIMOS line-strength indices\\
IMF mass factor relative to Kroupa & $\alpha_{\rm MW}$ & $1.04\pm0.15$  &   $M_*/L_{\rm F814W}$ and   $\Upsilon_{\rm MW}$  \\
\hline
\end{tabular}
\label{tab:summary}
\end{table*}

The previous sections have presented measurements or estimates for the total lensing mass $M_{\rm Ein}$, the dark matter mass $M_{\rm DM}$, the 
luminosity $L_{\rm F814W}$,  and the stellar population model mass-to-light ratio assuming a Kroupa IMF ($\Upsilon_{\rm MW}$). All of these quantities refer to 
mass and luminosity projected within the Einstein radius. Combining these inputs, the IMF mass normalisation factor is simply
\[
\alpha_{\rm MW} = \frac{M_{\rm Ein} - M_{\rm DM}}{L_{\rm F814W}}  \cdot \frac{1}{ \Upsilon_{\rm MW}}  \, .
\]
In practice of course, each quantity above is described by a probability distribution, which can be approximated as lognormal for $M_{\rm DM}$ and normal for the other
variables. Sampling from these distributions, we arrive at the probability distribution for $\alpha_{\rm MW}$, from which we 
determine whether various proposed IMFs are compatible with the observations for \egx. 
In this section, we first present the results using our preferred input parameters, and then explore the sensitivity of our result to various changes in the assumptions.

For the default result of this paper, we adopt 
	{\rev total mass $M_{\rm Ein}$\,=\,1.50$\pm$0.06$\times10^{11} M_\odot$ from lensing} (including systematic errors),
	dark matter mass $\log M_{\rm DM}$\,=\,10.34$\pm$0.25 from halo statistics,
	{\rev luminosity $L_{\rm F814W}$\,=\,4.07$\pm$0.08$\times10^{10} L_{\odot, \rm F814W}$\, from the HST photometry} (including 2\,per cent absolute calibration errors), and
	stellar-population mass-to-light ratio for Kroupa IMF $\Upsilon_{\rm MW}$\,=\,3.01$\pm$0.27$\,(M/L)_{\odot, \rm F814W}$  from the VIMOS spectrum fit (including errors from index systematics).
The adopted errors are conservative, including the various sources of systematic error assessed in earlier sections. Other possible systematics
are probed using robustness tests below.
With these inputs, the probability distribution for $\alpha_{\rm MW}$ is as shown in Figure~\ref{fig:alphapost} (thick black curve), and summarized in line 1 of Table~\ref{tab:alphapost}.
The distribution is fairly symmetric in $\alpha_{\rm MW}$, peaking at  $\alpha_{\rm MW}$\,=\,1.04, with a 68\,percent interval of $\pm$0.15. 
 Hence the results are consistent with a Milky-Way-like IMF, with either the Kroupa or the Chabrier form.
 A Salpeter or heavier IMF is disfavoured at the 99.8\,per cent confidence level\footnote{Note this is a one-tailed confidence limit, 
 i.e. for a Gaussian distribution 84.1\,per cent would correspond to a +1$\sigma$ deviation, and 99.8\,per cent would correspond to +2.9$\sigma$. 
 In practice, the distribution is not quite Gaussian, so we quote the probabilities estimated directly from the high-$\alpha_{\rm MW}$ tail.}, 
 while a heavyweight IMF with  $\alpha_{\rm MW}$\,$\ge$\,2 is excluded with high significance.

An equivalent calculation for the F475W band using $\Upsilon_{\rm MW}$\,=\,6.32$\pm$0.63$\,(M/L)_{\odot, \rm F475W}$
and the F475W luminosity of 2.07$\pm$0.04$\times10^{10}\,L_\odot$ yields
$\alpha_{\rm MW}$\,=\,0.97$\pm$0.15, consistent with the default result (thick green curve in Figure~\ref{fig:alphapost} and line 2 of Table~\ref{tab:alphapost}). 
This agreement simply confirms that the best-fitting stellar population model correctly predicts the observed F475W--F814W colour within $R_{\rm Ein}$.

We have seen that the choice of Balmer indices in the stellar population fitting has the largest impact on the derived $\Upsilon_{\rm MW}$. This propagates
trivially to the results for  $\alpha_{\rm MW}$ 
(blue curves in Figure~\ref{fig:alphapost} and lines 3--4 of Table~\ref{tab:alphapost}).
If only H$\beta$ is used for age constraints, the best $\alpha_{\rm MW}$ shifts upwards to 1.17,
and the confidence with which a heavier-than-Salpeter IMF is excluded is reduced to 99.4\,per cent. The heavyweight IMF remains firmly rejected.
Allowing for  $\alpha$-enhanced model populations being slightly brighter than solar-scaled abundance models (Percival et al. 2009), the derived $\alpha_{\rm MW}$
would be increased by $\sim$2\,per cent.

Our treatment of the dark matter contribution incorporates the expected intrinsic scatter among halos, under the assumption of pure dark-matter clustering.
In practice, the halo of \egx\ may deviate from the assumptions of this model, especially in the innermost regions, where the dark matter distribution
may contract in response to the dominant baryonic component. Simulations by different groups differ in their estimates of the strength of this effect (e.g. see discussion
in Gnedin et al. 2011). 
Reviewing comparisons of hydrodynamic simulations against dissipationless control simulations, we note the following:
Gnedin et al. (2011) find enhancements in the the inner dark-matter mass (enclosed within 1\,per cent of the halo virial radius) by factors of 2--4.
Johansson, Naab \& Ostriker (2012) find the central dark-matter mass (enclosed within 2\,kpc) is enhanced by a factor of 2.3 (their halo A2).
Remus et al. (2013) found central dark-matter density enhanced by a factor of 2--3 (their figure 1).
These results are generally for galaxy-scale halos. For $10^{13}$\,$M_\odot$ groups (more relevant to \egx\ / Abell S0740), Duffy et al (2010) 
find smaller enhancements, between zero and 50\,per cent in the inner dark-matter density, depending on the adopted feedback prescription.
All of these factors refer to three-dimensional densities or enclosed masses, rather than projected quantities.
On balance, we adopt a factor of two as an upper limit to the likely effect of halo contraction. Rescaling our input distribution of $M_{\rm DM}$ by this
factor, we would recover $\alpha_{\rm MW}$\,=\,$0.91^{+0.17}_{-0.21}$ (purple dashed curve in Figure~\ref{fig:alphapost} and line 5 of Table~\ref{tab:alphapost}).
An alternative limiting case is to assume that dark matter is negligible within the Einstein radius, so that stars must account for the entirety of the lensing mass
(red dashed curve in Figure~\ref{fig:alphapost} and line 4 of Table~\ref{tab:alphapost}).
Under this extreme model, the best $\alpha_{\rm MW}$ is $1.20^{+0.13}_{-0.11}$, which is marginally consistent with a heavier-than-Salpeter IMF (98\,per cent), but 
still incompatible with the heavyweight models.

We neglected the extra mass that would be contributed by a central super-massive black hole.
From the $M_{\rm BH}$--$\sigma$ relationship given by McConnell et al. (2011) for early-type galaxies, 
the mean expected black hole mass is $2.5\times10^9$\,$M_{\odot}$, or 2\,per cent of $M_{\rm Ein}$.
If we account also for the intrinsic scatter of 0.38\,dex around the  McConnell et al.  $M_{\rm BH}$--$\sigma$ relation, 
the derived $\alpha_{\rm MW}$ would be slightly reduced, relative to our default solution, to 1.01$\pm$0.15.

For the default result, we used the M05 stellar population models because these provide the most convenient predictions for both mass-to-light ratios
and line-strength indices in metal rich, $\alpha$ enhanced populations.
We have already noted that a full-spectrum fitting approach, using the CvD12a models, yields essentially identical $\Upsilon_{\rm MW}$, and consequently 
the same result for $\alpha_{\rm MW}$.
To test the effect of using other model sets, we discard all information from the 
line strength indices, and instead impose an external prior for the age. 
Given the mass of the galaxy, its pure absorption spectrum including absence 
of any emission at H$\alpha$ (from the 6dF spectrum) and smooth light distribution (even in the central regions, where dust features and star-forming rings
are sometimes seen in HST observations of ellipticals --- {\rev e.g. Laine et al. 2003; Martel et al. 2004}), it is unlikely that \egx\ has experienced substantial star-formation since $z$\,$<$\,1.
We adopt a Gaussian prior on (I-band luminosity-weighted) formation redshift with mean 2.5 and standard deviation 0.75.
Combining this with the M05 predictions, using {\sc EzGal}, for 1.5-times-solar metallicity (the maximum implemented for all model sets), we obtain a predicted stellar
 mass-to-light ratio 
	$\Upsilon$\,=\,$2.67^{+0.15}_{-0.25}\,(M/L)_{\odot, \rm F814W}$.
This is slightly smaller than our spectroscopic estimate, since the spectroscopy favours earlier formation and higher metallicity.
Combining this estimate with lensing and the dark-matter contributions would yield $\alpha_{\rm MW}$\,=\,$1.18_{-0.17}^{+0.16}$. 
We can now derive equivalent estimates for other stellar population models, using the same external age prior, and compare to this baseline value.
The results for $\alpha_{\rm MW}$ are:
0.96 for Bruzual \& Charlot (2003)\footnote{Unchanged if we use instead the unpublished Charlot \& Bruzual 2007 version.};
1.10 for Conroy, Gunn \& White (2009); 
1.05 for Percival et al. (2009);
1.12 for Fioc \& Rocca--Volmerange (1997), all with uncertainties of $\sim$0.15.
In all cases, the models are as implemented by default in {\sc EzGal} (see Mancone \& Gonzalez 2012), with the maximum 
1.5-times-solar metallicity. Hence for common assumptions on the galaxy age, other stellar population models yield {\it smaller} IMF 
normalisations than M05, 
by up to 20\,per cent\footnote{It should not be assumed that identical shifts would apply to the {\it full} analysis including spectroscopy, 
since the other models may predict slightly different index strengths as well as different mass-to-light ratios.}.
We conclude that the derived IMF constraint is fairly insensitive to the choice of stellar population models among the currently favoured sets.

Finally, we note that rescaling the distance assumed for \egx\ affects $\alpha_{\rm MW}$ linearly. If instead of placing the galaxy at its Hubble-flow distance, we assign
it the same distance as Abell 3570, then $\alpha_{\rm MW}$  is increased to 1.15. If instead we allow the galaxy a large positive peculiar velocity\footnote{\egx\
lies in the foreground of the Shapley supercluster; the original ACS data were obtained as part of an effort to measure peculiar velocities in this region.}
of 1000\,\kms, then $\alpha_{\rm MW}$  is reduced to 0.95.

To summarize, using a combination of lensing and stellar population constraints, with correction for dark matter contributions based on simulations, 
we find that \egx\ has a stellar mass-to-light ratio compatible with a Milky-Way-like (Kroupa or Chabrier) IMF. 
{\rev A Salpeter IMF is significantly disfavoured, and a heavyweight IMF is excluded.}
The statistical errors in the IMF mass normalisation factor are $\sim$15\,per cent; a range of robustness tests suggest that systematic errors are also 10--15\,per cent. 
The most relevant measured and derived parameters for the \egx\ system are provided in Table~\ref{tab:summary} for reference.

\section{Discussion}\label{sec:disc}

\begin{figure}
\includegraphics[angle=0,width=85mm]{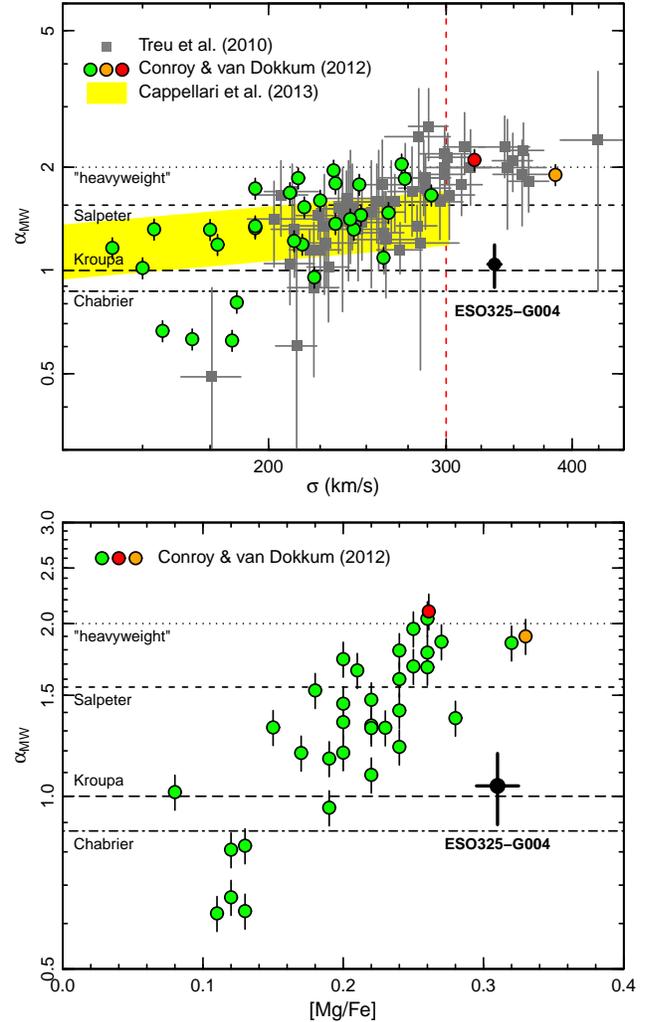}
\caption{
Upper panel: result for \egx\ compared to the $\alpha_{\rm MW}$--$\sigma$ relations from the SLACS lenses (T10), 
spectroscopic analysis (CvD12b) and stellar dynamics (Cappellari et al 2013). We highlight the $\sigma$\,$>$\,300\,\kms\ 
regime probed by \egx, in which both T10 and CvD12b favour heavyweight IMFs on average.
Lower panel: equivalent comparison for the $\alpha_{\rm MW}$--[Mg/Fe] relation from CvD12b. The vertical axis scale is 
not identical to the upper panel.
Velocity and [Mg/Fe] are as measured within an aperture of $R_{\rm eff}/8$ for \egx\ and CvD12b, and corrected to this 
aperture for T10 and Cappellari et al. (2013). In both figures, the red point represents a stacked spectrum of 
four massive ellipticals in Virgo from van Dokkum \& Conroy (2010). The orange point is M87.
}
\label{fig:alphacomp}
\end{figure}

In this section, we examine how the results obtained for \egx\ compare to recent results which favour heavier IMFs in ellipticals with similar properties. 
We focus in particular on the apparent disagreement between our results and those of T10 and CvD12b, and speculate on potential explanations
for this tension.

\subsection{Comparison to SLACS lensing results}

As described in Section 1, SLACS is a spectroscopically-defined lens sample based on SDSS. Lens systems were identified from 
the presence of discordant emission lines in the spectra. For the main study, HST follow-up was obtained for systems with at least two lines,
in practice usually [O\,{\sc ii}]\,3727\,\AA\ plus H$\beta$ or  [O\,{\sc iii}]\,5007\,\AA. A few additional targets with strong  [O\,{\sc ii}] detections, 
but no corroborating lines, were also followed up. 

Figure~\ref{fig:alphacomp} (upper panel) compares our result for \egx\ against the correlation of $\alpha_{\rm MW}$ with velocity dispersion ($\sigma$) 
from SLACS, among other works.
We have increased the velocity dispersions from T10 by 7\,per cent, as an aperture correction to $R_{\rm eff}/8$ (this
aperture is selected for consistency with CvD12b in the same figure, discussed below). Apart from this correction, and conversion to our definition 
of $\alpha_{\rm MW}$, the points are as in figure 4 of T10. 
As reported by T10, the SLACS lenses follow a clear trend of increasing IMF mass normalization with increasing $\sigma$, 
reaching $\alpha$\,$\approx$\,2 at $\sigma$\,$>$\,300\,\kms. As they also note, the observed scatter is compatible with 
{\it no} intrinsic dispersion around the $\alpha_{\rm MW}$--$\sigma$ relation.
The twelve galaxies with $\sigma$\,$>$\,300\,\kms\ (after aperture correction to $R_{\rm eff}/8$ as in Figure~\ref{fig:alphacomp})
have a mean  $\alpha_{\rm MW}$ of 2.04, and a $\chi^2$ of only $2.3$ around this 
mean (Pr($\chi^2_{\nu=11}$\,$\leq$\,2.3)\,=\,0.003).
At face value, then, our measurement of $\alpha_{\rm MW}$\,=\,1.04$\pm$0.15 for \egx, a galaxy with a similar velocity dispersion, 
is not only significantly different from the {\it mean} SLACS $\alpha_{\rm MW}$, but 
is also inconsistent with the {\it distribution} of $\alpha_{\rm MW}$  from SLACS at comparable velocity dispersion.
The small lensing mass of \egx\ shows, at the very least, that not all $\sigma$\,$>$\,300\,\kms\ ellipticals have heavyweight IMFs,
contrary to the implications of T10. 

The SLACS analysis method differs in several ways from our approach, for example using only colour information to derive the age and metallicity, 
rather than high-S/N spectroscopy, and fitting a halo model directly to each lens using the measured velocity dispersion of the galaxy, instead of 
using simulation statistics. Additionally, the lensing geometry of \egx\ is quite atypical of the average properties of SLACS sample lenses, 
due to differences in selection/discovery methods.

The \egx\ lens system differs from SLACS systems in both the redshift of the lens and that of the source. The redshift of \egx\  is smaller than
that of any SLACS lens, and a factor of 7.5 smaller than the median for the $\sigma$\,$>$\,300\,\kms\ SLACS lenses.
Assuming only that dark matter follows a more extended profile than the stellar mass, it follows that the dark-matter fraction
within the Einstein radius is an increasing function of the ratio $R_{\rm Ein}/R_{\rm eff}$. 
For \egx, this ratio is 0.23, compared to a median of 0.62 for  the twelve $\sigma$\,$>$\,300\,\kms\ SLACS lenses (Auger et al. 2009).
Hence in \egx, the dark-matter contribution is smaller, and all of the systematic and random 
uncertainties in modelling the dark matter are suppressed\footnote{Note that T10 report average dark-matter fractions of
only $\sim$20\,per cent, but this is contingent upon a particular (spherical uncontracted NFW) model for the halos.}.

\begin{figure}
\includegraphics[angle=0,width=85mm]{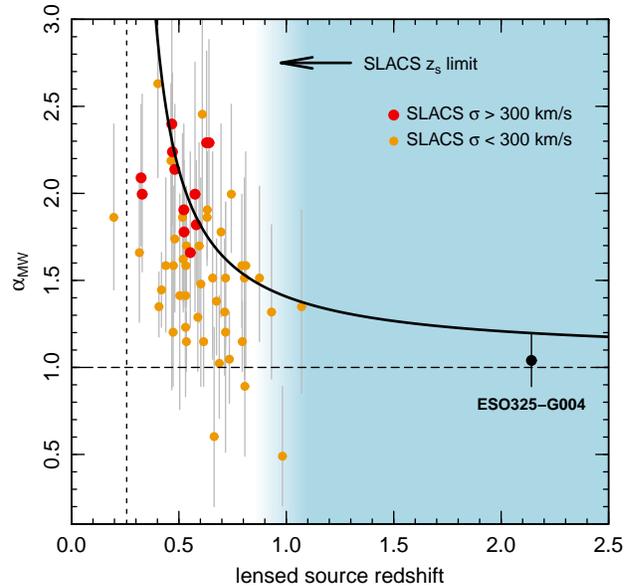}
\caption{
Derived $\alpha_{\rm MW}$ versus source galaxy redshift for SLACS and for \egx. 
Red points highlight the most massive SLACS lenses ($\sigma$\,$>$\,300\,\kms). 
The black curve shows the relationship between $\alpha_{\rm MW}$ and source
redshift through the lensing geometry factor ($f_z$\,=\,$D_{\rm s} / D_{\rm ls}$), computed for the median lens redshift 
$\langle$$z_{\rm l}$$\rangle$\,=\,0.26
of the massive SLACS galaxies (shown as the vertical dashed line),
and normalized  to $\langle$$\alpha_{\rm MW}$$\rangle$\,=\,2.04 found for these galaxies. 
{\rev The curve indicates the implied $\alpha_{\rm MW}$ if these galaxies were lensing sources at other redshifts. The
source redshift limit imposed by the SDSS spectroscopic selection translates to a limit of  $\alpha_{\rm MW}$\,$\ga$\,1.5 for the
high-$\sigma$ SLACS lenses. 
Note that the curve depends on $z_{\rm lens}$. The 
lower-$\sigma$ SLACS systems (orange)  have smaller $z_{\rm lens}$ on average, so these galaxies can lens low-redshift sources
even for $\alpha_{\rm MW}$\,$\la$\,1.5.
}
}
\label{fig:slacsbias}
\end{figure}

At $z_{\rm_s}$\,=\,2.141, the redshift of the source in \egx\ is much larger than that in any SLACS system. The absence of high-redshift sources in SLACS 
is a simple consequence of the spectroscopic selection method: for $z_{\rm_s}$\,$>$\,0.8--0.9, the H$\beta$ and [O\,{\sc iii}] lines shift out of the SDSS spectral
range; the single-line [O\,{\sc ii}] objects provide a few higher redshift lenses, but there are none at $z_{\rm_s}$\,$>$\,1.1
For a given lens redshift and Einstein radius, a closer source implies a more massive lens galaxy, so the SLACS 
selection of low-redshift source galaxies potentially biases their sample towards lenses with large central (stellar plus dark matter) 
masses.
This effect could be the cause of the significant ($\sim$3$\sigma$) anti-correlation between
$\alpha_{\rm MW}$  and source redshift in the SLACS sample (Figure~\ref{fig:slacsbias}).
For the twelve  $\sigma$\,$>$\,300\,\kms\ SLACS lenses, 
the mean source redshift is 0.52, and the mean lens redshift is 0.26. 
{\rev If we assume that  $\sigma$\,$>$\,300\,\kms\ galaxies actually span a wide range in $\alpha_{\rm MW}$, we can ask which values correspond
to source redshifts that are detectable in SDSS. Computing the geometric factor for $z_{\rm_l}$\,=\,0.26 and a range of $z_{\rm_s}$, we find that the  
source-redshift selection limit of $z$\,$<$\,0.85 (for multiple-line source detection) imposes a limit of $\alpha_{\rm MW}$\,$>$\,1.5
for these galaxies.}
Sources at  $z$\,$\ga$\,2,  would correspond to $\alpha_{\rm MW}$\,$\approx$\,1.2, but are
undiscoverable with the SLACS approach.
{\rev Hence SLACS might be selecting only the highest-$\alpha_{\rm MW}$ galaxies within a broad intrinsic distribution.}
{\rev A counter-argument to this suggestion is that Auger et al. (2010) find the SLACS lens galaxies to follow the same Fundamental Plane correlations as derived 
for general SDSS samples, which would appear to argue against such a bias, unless either the distribution {\it within} the plane was unrepresentative.}
A full investigation of source-redshift selection bias is beyond the scope of the present paper, but could likely be carried
out by generalizing the methods of Arneson, Brownstein \& Bolton (2012).

By contrast, our serendipitous morphological discovery of arcs behind \egx, and subsequent 
spectroscopy to secure the source redshift, is free from this ``source-redshift'' bias. 
Hence, if there is a very broad intrinsic distribution of $\alpha_{\rm MW}$
(or alternatively, a distribution in deviations from the assumed dark-matter halo properties), SLACS may select only 
those galaxies at the massive extreme, while \egx\ provides a single but {\rev more representative} sample from the distribution. 
In this context we note that Spiniello et al. (2011), analysing a morphologically-identified $\sigma$\,$\approx$\,340\,\kms\ lens 
with $z_{\rm s}$\,=\,2.38 (Belokurov et al. 2007), rule out very heavy IMFs, while Milky-Way-like or Salpeter distributions are 
compatible with lensing and dynamical constraints. 
{\rev 
This result broadly supports our suggestion that different methods of selecting lenses may lead to different distributions for the recovered $\alpha_{\rm MW}$; further
follow-up of large morphologically-defined lens samples is required to test this possibility.
}

\subsection{Comparison to CvD dwarf-star-indicator method}

The spectroscopic method used by CvD12b is not sensitive explicitly to mass, but instead to the characteristic features
of dwarf stars in the integrated spectra of galaxies. The implications for $\alpha_{\rm MW}$ are derived assuming a three-part
power-law IMF, fixed to the Salpeter slope at $M$\,$>$\,$M_\odot$ but allowed to vary to steeper or shallower slopes at lower mass.
CvD12b find an increasing trend of $\alpha_{\rm MW}$ as a function of velocity dispersion, and also as a function of the Mg/Fe
abundance ratio\footnote{These two properties are correlated. 
A fit for $\alpha_{\rm MW}$ versus both $\sigma$ and [Mg/Fe] suggests the latter is dominant in driving the relationship.}.
The CvD12b sample has few galaxies at the high velocity dispersions and Mg/Fe ratios similar to \egx.
At $\sigma$\,$>$\,300\,\kms, the sample contains only M87 and a stacked spectrum of four Virgo cluster 
galaxies from the original van Dokkum \& Conroy (2010) study. These spectra both yield $\alpha_{\rm MW}$\,$\approx$\,2,
in agreement with the SLACS results at similar velocity dispersion. 
La Barbera et al. (2013) have analysed dwarf-sensitive features in a stacked sample of early-type galaxies from SDSS, 
assuming single or broken power-law IMFs. As in CvD12b, there is a strong trend of $\alpha_{\rm MW}$ with velocity dispersion.
 At $\sigma$\,$\approx$\,300\,\kms, the best-fit broken-power-law IMF 
models have $\alpha_{\rm MW}$\,$\approx$\,1.7, 
similar to the CvD12b results at similar $\sigma$. 
Spiniello et al. (2013), using a different set of spectral features, recover a weaker
dependence of IMF slope versus $\sigma$, with slopes only mildly steeper than Salpeter at 
$\sigma$\,$\approx$\,300\,\kms.

Figure~\ref{fig:alphacomp} compares \egx\ with the CvD12b trends in both the $\alpha_{\rm MW}$--$\sigma$ and the 
$\alpha_{\rm MW}$--[Mg/Fe] relations. For consistency with CvD12b in placing \egx\ on the horizontal axis of the figures, we
use the velocity dispersion and [Mg/Fe] ratio measured from a spectrum extracted from the VIMOS data within a radius of $R_{\rm eff}/8$. 
The abundance ratio is derived using full-spectrum-fitting to the CvD12a models, allowing for variation in other relevant parameters 
(age, Fe/H, C/Fe). This analysis yields [Mg/Fe]\,=\,0.31$\pm$0.02 and $\sigma$\,=\,331$\pm$2\,\kms, 
where errors are derived from repeatability over the separate VIMOS exposures.

The lensing-derived $\alpha_{\rm MW}$ for \egx\ is inconsistent with the 
{\it average} $\alpha_{\rm MW}$ from CvD12b, for galaxies of similar properties. 
{\rev However, the intrinsic scatter
at high $\sigma$ is poorly determined, so this discrepancy could simply indicate that \egx\ has a lighter-than-average IMF within a broad underlying distribution.
In this scenario, we would expect that this galaxy would also exhibit weaker dwarf-star signatures than average for massive galaxies.}

\begin{figure}
\includegraphics[angle=0,width=85mm]{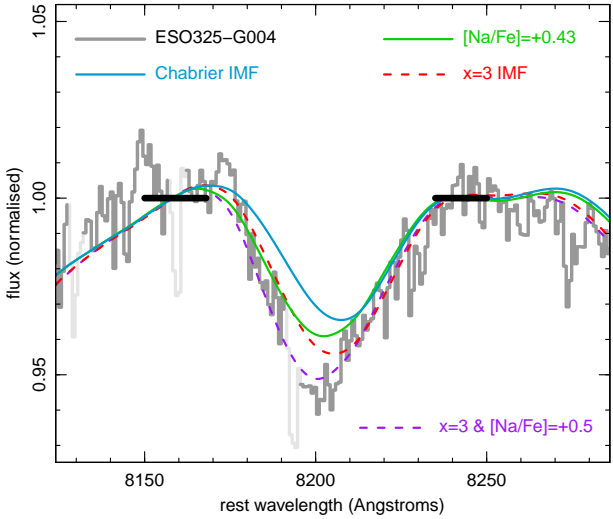}
\caption{
The dwarf-sensitive Na\,{\sc i} feature in \egx, compared to models from CvD12a. 
Light grey sections indicate pixels affected by strong sky lines.
All models shown have [$\alpha$/Fe]\,=\,+0.3 and age 13.5\,Gyr. 
The spectra are normalized at the continuum regions of the index defined by Spiniello et al. (2012), shown as black bars.
The blue and green curves are Chabrier IMF models with solar and enhanced sodium abundances, respectively.  
[Na/Fe]\,=\,+0.43 is the average obtained for $\sigma$\,$\approx$\,300\,\kms\ galaxies in SDSS by Conroy et al. (2013a). 
A bottom-heavy IMF model, with $x$\,=\,3 (in the convention where the Salpeter slope is 2.3)
is shown in red. Matching the observed spectrum requires both $x$\,=\,3, {\it and} enhancement of sodium (purple line). 
The $x$\,=\,3 IMF models would have $\alpha_{\rm MW}$\,$>$\,2, and hence would violate the lensing limits. 
}
\label{fig:forsnai}
\end{figure}

{\rev To test this possibility, we have obtained a far-red spectrum of  \egx\, to measure the \nai\ feature and hence compare dwarf-star indicators 
versus lensing constraints directly.}
We observed \egx\ with FORS2 (Appenzeller et al. 1998) at UT1 of the VLT, on 2013 May 10.
The observations were made using the 1028z grism, with a 1.3\,arcsec slit width, providing 
a wavelength coverage of 7730--9500\,\AA, with 2.5\,\AA\ FWHM resolution, sampled at 0.8\,\AA\ per pixel.
The total exposure time was 1570\,s.
To mimic a circular aperture measurement sampling the light within $R_{\rm Ein}$, we extracted the spectrum within  $\pm$2.85\,arcsec from the galaxy centre, weighted 
linearly with distance. 
Figure~\ref{fig:forsnai} shows the Na\,{\sc i} region in the resulting spectrum (which has $S/N$\,$\approx$\,150\,\AA$^{-1}$), in comparison with models from 
Conroy \& van Dokkum (2012a)\footnote{Strictly, the models shown are updated versions, with improved abundance response functions.}.
The strength of  Na\,{\sc i} in \egx\ appears similar to that in other high-$\sigma$ elliptical galaxies (van Dokkum \& Conroy 2012; Conroy \& van Dokkum 2012b; Ferreras et al. 2013). 
The observed absorption is much stronger than in the models with Milky-Way-like IMF, even allowing for enhancement
of sodium abundances by 0.43\,dex (the average for $\sigma$\,$\approx$\,300\,\kms\ found by Conroy, Graves \& van Dokkum 2013).
To reproduce the observed Na\,{\sc i} feature in the CvD12a models would require either a substantially steeper IMF (e.g. a single 
power law with slope $x$\,$\ga$\,3 in the 
convention where Salpeter has $x$\,=\,2.35) or a larger enhancement in sodium, or some combination of these effects. 
For the Chabrier IMF, which is consistent with the lensing constraint, the sodium enhancement would
have to be quite extreme, e.g. [Na/Fe]\,$\approx$\,+1.2, {\rev which is not supported by the strength of the Na\,D absorption in the VIMOS spectrum.}.

Although a single power law with $x$\,$\ga$\,3 would certainly violate the lensing mass constraint, it is conceivable that a more flexible IMF prescription 
as used by CvD12b would be able to reproduce the observed \nai\ without requiring excessive mass contributions from low-mass stars. If so, it
may be possible to use the lensing mass in combination with the spectroscopic signatures to probe the detailed shape of the IMF. 
For example La Barbera et al. (2013) have shown a comparison of $M_*/L$ derived from dwarf-star indicators (including \nai)
against dynamical estimates, which excludes single power laws, but yields consistent results when two-part broken power-law
IMFs are adopted.  
\egx\ provides an opportunity to conduct a similar test for an individual galaxy using a robust external mass 
estimate\footnote{\rev While this paper was under review, Barnab\`e et al. (2013) published an analysis along these lines, constraining
the slope and low-mass cut-off for a power-law IMF in two SLACS lenses.}.

{\rev In summary, our lensing measurement of $\alpha_{\rm MW}$ for \egx\ is inconsistent with the  average derived from CvD12b for galaxies of similar properties. 
While it is possible that \egx\ is an outlier from a distribution of $\alpha_{\rm MW}$, the strong measured \nai\ absorption does not support this interpretation, unless the 
sodium abundance is much larger than average for $\sigma$\,$\approx$\,300\,\kms ellipticals. 
The \nai\ measurement suggests some tension between the lensing mass and the IMF-sensitive spectral features for this galaxy, 
but further work is required before a firm conclusion can be drawn.}

\subsection{Comparison to stellar dynamics}

{\rev 
We comment here briefly on comparison to recent dynamical estimates of $M_*/L$ in early-type galaxy samples, and 
make a dynamical estimate for the mass of \egx. 
}

The upper panel of Figure~\ref{fig:alphacomp} includes the estimated $\alpha_{\rm MW}$--$\sigma$ relation from Cappellari et al. (2013) derived from the Atlas3D survey 
(from their figure 13, upper panel).
The velocity dispersions have been increased by 10\,per cent as an approximate aperture correction to $R_{\rm eff}/8$ in common with
the other data sources plotted. 
Cappellari et al. recover a much shallower $\alpha_{\rm MW}$--$\sigma$ relation than T10, but it should be noted 
that Atlas3D and SLACS overlap only for $\sigma$\,=\,200--300\,\kms, and in this interval the agreement is fairly close. 
\egx\ does not lie within the $\sigma$ range probed by Atlas3D. Extrapolation of the Cappellari et al. trend would suggest
a Salpeter-like $\alpha_{\rm MW}$\,$\approx$\,1.5. Allowing also for possible intrinsic scatter (estimated as 20\,per cent at lower $\sigma$), 
\egx\ is marginally consistent with the Cappellari et al. trend. 

A smaller dynamical study of ellipticals in the Coma and Abell 262 clusters (Thomas et al. 2011; Wegner et al. 2012) 
finds a trend which appears more similar to the SLACS trend, with average $\alpha_{\rm MW}$\,$\approx$\,2 for the five Coma galaxies with
$\sigma$\,$>$\,300\,\kms\ (after aperture correction to $R_{\rm eff}/8$).

{\rev
We can make a crude estimate of the dynamical mass of \egx\ using the virial mass estimator, $M_{\rm dyn}$\,=\,$5\sigma_{\rm Eff}^2\,R_{\rm Eff}/G$, 
where $\sigma_{\rm Eff}$ is the velocity dispersion estimated within the effective radius. 
Using the measured half-light radius of 8.5\,kpc\ and $\sigma_{\rm Eff}$\,=\,310\,\kms\ (allowing for an 8\,per cent aperture correction from $R_{\rm Ein}$ to $R_{\rm Eff}$), this
estimator yields  $M_{\rm dyn}$\,=\,9.4$\pm$1.3$\times$10$^{11}$\,$M_\odot$. The error is derived from the galaxy-to-galaxy scatter of 14\,per cent found by
Cappellari et al. 2006 through comparison to masses derived from Schwartzschild models. 
This quantity should represent the total mass extrapolated to large radius.
From the luminosity profile, we find that 18\,per cent of the total flux is projected inside the Einstein radius, so for constant mass-to-light ratio, 
the {\it dynamical} estimate of $M_{\rm Ein}$ is 1.7$\pm$0.2$\times10^{11}\,M_\odot$, which is consistent with the lensing estimate. 
Note that $M_{\rm Ein}$ from lensing is a much more direct and robust measurement of the mass enclosed within a small aperture, where dark-matter contributions are small and 
the stellar populations well determined. The value of $\alpha_{\rm MW}$ from lensing should thus be more reliable and accurate than dynamical estimates. 
}

\subsection{A possible correlation with compactness?}\label{sec:compactness}

{\rev

While this paper was under revision, L\"asker et al. (2013) reported detailed dynamical models for an unusually-compact elliptical (``b19'') with $\sigma$\,$\approx$\,360\,\kms, which 
appear to require a very heavy IMF ($\alpha$\,$\sim$\,2). The striking contrast between \egx\ and b19 (which has similar velocity dispersion but  $\sim$7 times higher luminosity
surface density) is suggestive of a possible correlation of $\alpha_{\rm MW}$ with galaxy compactness.

Also during revision of our paper, Conroy et al. (2013b) published average $\alpha_{\rm MW}$ from dwarf-star indicators and dynamical estimates,
for a sample of compact elliptical galaxies in SDSS. Their sample definition selects the $\sim$6\,per cent densest early-type galaxies, based
on stellar mass surface density. For such galaxies the contribution of dark matter within the SDSS fibre is sufficiently small to justify assuming mass follows light in the dynamical models.
A consistent trend of increasing $\alpha_{\rm MW}$ with increasing $\sigma$ is recovered using both methods, but both dynamics and 
spectral features yield systematically larger $\alpha_{\rm MW}$ for compact ellipticals than found for the CvD12b sample. 
For example in the highest-$\sigma$ bin, with $\sigma$\,$\approx$\,300\,\kms, 
the dwarf-star-indicators suggest $\langle\alpha_{\rm MW}\rangle$\,$\approx$\,2.3, compared to $\sim$1.7 at the same velocity dispersion in CvD12b. 

Both of these recent advances imply that the IMF may vary among galaxies of similar velocity dispersion, in a way that is correlated with galaxy compactness, 
and hence presumably to the degree of dissipation in the early formation history. Accounting for this modulation may eventually help reconcile \egx\ with results from other studies.

}

\section{Conclusions}\label{sec:concs}

In this paper, we have presented new data on the \egx\ lens system, which demonstrate that {\it this} giant elliptical galaxy does {\it not} 
have a very heavy IMF of the type suggested for similar galaxies in several recent works, in particular T10 and CvD12a. 
The IMF mass normalisation, relative to the Milky Way (Kroupa) is $\alpha_{\rm MW}$\,=\,1.04$\pm$0.15, consistent with either 
a Chabrier or a Kroupa IMF. An IMF heavier than Salpeter ($\alpha_{\rm MW}$\,$>$\,1.55) is disfavoured at the $>$99.8\,per cent level.
This result is robust against a range of possible systematic errors, and to the treatment of dark matter contributions. Even if 
we attribute all of the lensing mass to stars, the IMF is lighter than Salpeter at the 98\,per cent confidence level. 

One explanation for the difference between our result and those favouring heavyweight IMFs is simply that \egx\ is
unusual among massive ellipticals: i.e. on average such galaxies have $\alpha_{\rm MW}$\,$\approx$\,2, but there is some
intrinsic scatter around this value, and the closest-known strong lensing elliptical happens to have a much
lighter IMF. 
This possibility cannot be excluded, but we have highlighted two lines of evidence to the contrary.  
First, the high-$\sigma$ SLACS lenses {\it all} have $\alpha_{\rm MW}$\,$\approx$\,2, apparently
with {\it no} intrinsic scatter (and indeed an observed scatter 
almost too small to be consistent with the errors). This suggests that the \egx\ differs systematically 
from the SLACS sample lenses, perhaps due to the different selection/discovery methods involved.
Second, the strong Na\,{\sc i} feature observed in \egx\ suggests that this galaxy is similar to other 
massive ellipticals in having enhanced dwarf-sensitive spectral features.
As we have indicated, a full analysis of these features is required before firm conclusions
can be drawn, but since a Milky-Way-like IMF does not match the \nai\ data without extreme Na/Fe ratios, 
there appears to be some tension between the two methods for this galaxy. 
{\rev We have noted the very recent hints that compact galaxies have heavier IMFs (at given $\sigma$) than normal ellipticals, which
could help to resolve some of these apparent disagreements.}

Observations of a single galaxy cannot provide a definitive answer as to whether massive ellipticals, as a class,
formed their stars according to a mass distribution different from that in the Milky Way. 
Nevertheless, the unique properties of \egx, the nearest-known strong-lensing galaxy, provide an important 
opportunity to inter-compare and calibrate the various methods of constraining the IMF. 
In future work, we intend to exploit further the the dwarf-star indicators measurable both from existing data and 
from infra-red spectroscopy with KMOS (Sharples et al. 2013). We will also use the IFU data to build dynamical models, subject 
to the lensing constraints, as a further probe of the mass distribution in \egx.

\section*{Acknowledgements}
RJS was supported by STFC Rolling Grant PP/C501568/1 ``Extragalactic Astronomy and Cosmology at Durham 2008--2013''.
We thank ESO for the award of director's discretionary time for the \nai\ observations,
and Charlie Conroy for providing updated stellar population models in advance of publication. 
{\rev We thank the referee for several helpful comments and suggestions.}
This research has made use of the NASA/IPAC Extragalactic Database (NED) which is operated by the 
Jet Propulsion Laboratory, California Institute of Technology, under contract with the National Aeronautics and Space Administration.
 The Millennium Simulation databases used in this paper and the web application providing access to them were constructed as part 
 of the activities of the German Astrophysical Virtual Observatory.

{}

\label{lastpage}

\end{document}